\documentstyle[aps,preprint,prb,epsf]{revtex}
\begin{document}
\draft

\title{The Effects of $d_{x^2-y^2}$-$d_{xy}$ Mixing on Vortex Structures
and Magnetization}
\author{Mahn-Soo Choi and Sung-Ik Lee}
\address{Department of Physics, Pohang University of Science and
Technology, Pohang 790-784 South Korea}
\date{\today}
\maketitle

\begin{abstract}
The structure of an isolated single vortex and the vortex lattice, and
the magnetization in a $d$-wave superconductor are investigated within
a phenomenological Ginzburg-Landau (GL) model including the mixture of
the $d_{x^2-y^2}$-wave and $d_{xy}$-wave symmetry.  The isolated
single vortex structure in a week magnetic field is studied both
numerically and asymptotically.  Near the upper critical field
$H_{c2}$, the vortex lattice structure and the magnetization are
calculated analytically.    
\end{abstract} 


\section{Introduction}

Since the discovery of the high temperature superconductors (HTSC's),
a great number of experimental and theoretical investigations have
been carried out to identify the symmetry of the superconducting
pairing state.  Although there still remain controversies, it is
believed that the $d_{x^2-y^2}$-wave symmetry is most probable in
HTSC's; e.g. a series of strong evidences have been provided by the
phase sensitive experiments\cite{VanHar95}.   

In recent years, the structure of vortices in a $d_{x^2-y^2}$
superconductor was of great interest.  It was expected that the
structure of $d$-wave vortices might be very different from that of
$s$-wave vortices.  And connected with the vortex structure, the
question of order parameter symmetry admixture arose.

From a standpoint of the spontaneous symmetry breaking, in a bulk with
a perfect crystal symmetry, the order parameter transforms according
to an irreducible representation\cite{endnote11} of the
crystallographic point group\cite{Annett90a,Annett90b,Sigris91}.  Thus
in a crystal of tetragonal ($D_4$) symmetry, only one symmetry state
(e.g. $d_{x^2-y^2}$-wave) is allowed.  However near interfaces,
surfaces or impurities, the crystalline symmetry is not perfect and
the $d_{x^2-y^2}$-wave order parameter fluctuates spatially and hence
induces components of other symmetry.  This is also the case in the
presence of vortices, to which we will confine ourselves in this work.

Soininen {\it et al.}\cite{Soinin94}, starting from a simple
microscopic model Hamiltonian and its Bogoliubov-de Gennes
equation, found a substantial admixture of an induced $s$-wave and the
dominant $d_{x^2-y^2}$-wave. Based on this work, Berlinsky {\it et
al.}\cite{Berlin95} and Franz {\it et al.}\cite{Franzx96} investigated
the structure of $d$-wave vortices using the Ginzburg-Landau(GL)
theory.  Ren {\it et al.}\cite{Renxxx95} microscopically derived the
GL equation from the Gorkov equation of a continuum
mean field model, and it was studied in detail by Xu {\it et
al.}\cite{Xuxxxx95,Xuxxxx96} for the $d$-wave vortex structures.
Joynt\cite{Joyntx90} also included the $d$-$s$ mixing term in their
phenomenological GL model, from the symmetry argument.  Very recently,
Maki and B\'{e}al-Monod\cite{Makixx95} and Won and Maki\cite{Wonxxx95}
also incorporated $d$-$s$ mixture.  They used an interaction potential
with the $s$ channel repulsive Coulomb interaction in their
weak-coupling model.  In most of the works above, they observed that
the $s$-wave component induced near the vortex core causes the
$d$-wave component to have fourfold
anisotropy,\cite{Berlin95,Franzx96,Xuxxxx95,Xuxxxx96} and that the
vortices arrange in an oblique lattice instead of triangular
one.\cite{Berlin95,Franzx96,Xuxxxx96,Wonxxx95}

Recently, the possibility of admixture of $d_{x^2-y^2}$-wave and
$d_{xy}$-wave was also suggested.  Koyama and Tachiki\cite{Koyama96}
took into account the internal orbital motion of the pairing electrons
in their linearized Gorkov-type gap equation near $T_c$.  They found
that the dominant $d_{x^2-y^2}$-wave symmetry was mixed with
$d_{xy}$-wave symmetry, instead of the $s$-wave.   The coupling of the
two $d$-wave components led to a significant paramagnetic effect and
to strong enhancement of the upper critical field at lower
temperatures, which remarkably fits the experimental results on
over-doped cuprate superconductors Bi$_2$Sr$_2$CuO$_{y}$ and
Tl$_2$Ba$_2$CuO$_{6}$.  And Ichioka {\it et al.}\cite{Ichiok96}
considered both $d_{x^2-y^2}$-$s$ mixing and $d_{x^2-y^2}$-$d_{xy}$
mixing in the frame work of the classical Eilenberger equations.  They
studied the isolated single vortex structures and found that the
amplitude of the $d_{xy}$-wave component has the shape of an octofoil.

As described above, for the case of $d$-$s$ mixture, the single vortex
and/or vortex lattice structure have been extensively studied
\cite{Berlin95,Franzx96,Wonxxx95}, but not for the case of
$d_{x^2-y^2}$-$d_{xy}$ mixture.  In this paper, we present analogous
studies based on the GL theory which includes the mixing of
$d_{x^2-y^2}$-wave and $d_{xy}$-wave, assuming the $d_{x^2-y^2}$-wave
symmetry of the superconducting ground state.

\section{GL Theory for a $d$-Wave Superconductor}

We start with the GL free energy functional proposed by Koyama and
Tachiki:
\begin{eqnarray}
G(H) 
& = &  \int(dx)\;\Biggl\{ 
  \frac{\hbar^2}{4m_+}\left|
    \left(-i\nabla+\frac{2\pi}{\phi_o}{\bf A}\right) \Psi_+
  \right|^2
  + \alpha_+(T)|\Psi_+|^2 
  + \frac{1}{2}\beta_+|\Psi_+|^4  
  \nonumber \\
& & \mbox{}+ \frac{\hbar^2}{4m_-}\left|
    \left(-i\nabla+\frac{2\pi}{\phi_o}{\bf A}\right) \Psi_-
  \right|^2
  + \alpha_-|\Psi_-|^2 
  + \frac{1}{2}\beta_-|\Psi_-|^4 
  \nonumber \\
& & \mbox{}
  + \beta_X|\Psi_+|^2|\Psi_-|^2
  + \beta_Y|(\Psi_+^*\Psi_-^2 + h.c.)
  - \frac{1}{2}\gamma_p\left(
    \Psi_+^*\Psi_- + h.c.
  \right) B_z
  \nonumber \\
& & \mbox{}+ \frac{B^2}{8\pi} - \frac{{\bf B}\cdot{\bf H}}{4\pi}
\Biggr\},
\label{eq:GLmodel}
\end{eqnarray}
where the order parameter $\Psi_+$ ($\Psi_-$) corresponds to the
$d_{x^2-y^2}$-wave ($d_{xy}$-wave) symmetry.  As we previously assumed, in
Meissner state, the only non-vanishing component is $\psi_+$ and
$\alpha_+(T) < 0$ at $T < T_c$.   Other coefficients are assumed to be
positive and independent of temperature.

In passing, we make a couple of remarks.  In Eq. (\ref{eq:GLmodel}) we
included the coupling term $\sim(\Psi_+^{*2}\Psi_-^2+h.c.)$, which
was omitted in the model originally proposed by Koyama and
Tachiki\cite{Koyama96,endnote33}.  This term is necessary in order to get the most
general (up to the fourth order) free energy functional from the
symmetry consideration.\cite{Sigris91}  The mixing term
$\sim(\Psi_+^*\Psi_-+h.c.)B_z$ in the quadratic order give rise to the
paramagnetic current.\cite{endnote51}  As we will see below, it
dominates other mixing terms and significantly affects the mixed state
properties of $d$-wave superconductors.

In mean field approximation, neglecting thermal fluctuation effects,
physical properties are described by the corresponding GL equations.
It is convenient to introduce the dimensionless quantities by adopting
the fundamental length scale to be the GL coherence length $\xi_+$ of
$\Psi_+$ (See Table. \ref{table:parameters}).  In this case, the GL
equations and Maxwell equations are written as
\begin{equation}
\Pi^2 \psi_+ -  \psi_+ + |\psi_+|^2 \psi_+
+ \chi_+|\psi_-|^2\psi_+ + \zeta_+\psi_+^*\psi_-^2
- \nu_+B_z\psi_- = 0,
\label{eq:GL1}
\end{equation}
\begin{equation}
\Pi^2\psi_- + \Xi \psi_- + \Upsilon|\psi_-|^2\psi_-
+ \chi_-|\psi_+|^2\psi_- + \zeta_-\psi_-^*\psi_+^2
- \nu_-B_z\psi_+ = 0,
\label{eq:GL2}
\end{equation}
\begin{eqnarray}
\kappa^2{\bf J} 
& = &  -\frac{1}{2}(\psi_+^*\Pi\psi_+ + h.c.)
  - \frac{1}{2}\mu^{-1}(\psi_-^*\Pi\psi_- + h.c)
  \nonumber \\
& & \mbox{}+ \frac{1}{2}\nu_+\;
  \nabla\times(\psi_+^*\psi_- + h.c.)\hat{z}, 
\label{eq:Maxwell2}
\end{eqnarray}
where $\nabla\times\nabla\times{\bf A} = {\bf J}$ and $ \Pi =
(-i\nabla + {\bf A})$ and other parameters are defined
in the Table \ref{table:parameters}.  In Eq. (\ref{eq:Maxwell2}), the
last term is the paramagnetic current contribution.

\section{Isolated Single Vortex Structure}

Consider an isolated single vortex near the lower critical field
$H_{c1}$\cite{endnote21}.  For the problem of isolated single vortex,
it would be convenient to decompose the order parameter into the form
$\psi_\pm = f_\pm\:e^{+i\varphi_\pm}$.  Since the $d_{xy}$-wave
component is induced through the direct coupling to the magnetic
field, in the low field near $H_{c1}$ its amplitude is expected to be
very small compared with $d_{x^2-y^2}$-wave component.  Thus just for
intuitive understanding, for a moment we neglect the effect of the
coupling in the fourth order ($\chi_\pm\simeq 0$, $\zeta_\pm\simeq
0$).  The effect will be considered below. 

Let's first look at the phase distributions associated with the
vortex.  It is obvious from the rotational symmetry of the GL
equations that the phases should have $\varphi(r,\theta)_\pm =
-\theta$ up to an additive constant\cite{endnote1}, which we take to
be zero; the two $d$-wave components have the same winding.  Thus, the
differential equations are rewritten, in terms of $f_\pm$ only, as 
\begin{equation}
\Pi^2 f_+ -  f_+ + f_+^3 - \nu_+Bf_- = 0
\label{eq:GL11}
\end{equation}
\begin{equation}
\Pi^2f_- + \Xi f_- + \Upsilon f_-^3 - \nu_-Bf_+ = 0
\label{eq:GL12}
\end{equation}
\begin{equation}
\kappa^2{\bf J} 
= -(f_+^2+\mu^{-1}f_-^2)\:(-1/r + A)\hat{\theta}
+ \nu_+\;\nabla\times(f_+f_-)\hat{z}, 
\label{eq:Maxwell12}
\end{equation}
where 
$
\Pi^2 
= -\frac{d^2}{dr^2} - \frac{1}{r}\frac{d}{dr} +
\left(-\frac{1}{r} + A\right)^2
$,
${\bf A} = A(r)\hat{\theta}$, 
${\bf B} = B(r)\hat{z}$, and 
${\bf J} = J(r)\hat{\theta}$.

Second, near the vortex center ($r\to 0$),
$B(r)$ and $f_\pm(r)$ have
the asymptotic behavior of the following form:  
\begin{eqnarray}
B(r) 
& = & B_0 + B_2 r^2 + {\cal O}[r^4] \\
f_+(r) 
& = & C_0 r\left(
    1 + C_2r^2 + {\cal O}[r^4]
    \right) \\
f_-(r) 
& = & D_0 r\left(
    1 + {\cal O}[r^4]
    \right)
\end{eqnarray}
where\cite{endnote31} 
\begin{eqnarray}
C_2 
& = & -\frac{1}{8}(1+B_0)\left[
    1 + \nu_p^2 \frac{B_0^2}{(\Xi-B_0)(1+B_0)}
\right] \\
D_0
& = & \nu_2\frac{B_0}{\Xi-B_0}\; C_0 \\
B_2
& = & -\frac{1}{2}\frac{C_0^2}{\kappa^2}\left[
    1 -2\nu_p^2\frac{B_0}{\Xi-B_0} 
    + \nu_p^2\left(\frac{B_0}{\Xi-B_0}\right)^2
\right]
\end{eqnarray}
Note that since $B_0 \sim 2/\kappa^2$, in the extreme type-II
superconductors ($\kappa\gg 1$) the asymptotic behaviors of $B(r)$ and
$f_+(r)$ deviates very little from those of the conventional $s$-wave
GL theory ($C_2 \simeq -(1+B_0)/8$, $B_2 \simeq -C_0^2/2\kappa^2$).

Next, consider the region far from the vortex center ($r \gg \kappa$).
Assuming the extreme type-II superconductor ($\kappa \gg 1$), we can
treat $f_+=1$ and neglect $|\nabla^2 f|_- \sim f_-/\kappa^2 \ll f_-$.
Then by taking the curl of Eq. (\ref{eq:Maxwell12}) we get the usual
London equation:
\begin{equation}
\kappa^2\nabla\times\nabla\times{\bf B} + {\bf B} 
= 2\pi\delta^2({\bf r}), 
\end{equation}
and from Eq. (\ref{eq:GL12})
\begin{equation}
f_+(r) \simeq \frac{\nu_-}{\Xi}B(r) \sim
\sqrt{\kappa/r}\:e^{-r/\kappa},\;\;\; (r\gg\kappa).
\end{equation}
Thus far outside the core, only the pure $d_{x^2-y^2}$-wave component 
remains.

Now we discuss the effect of the mixing terms in the fourth order,
{\it i.e.} with finite $\chi_\pm$ and $\zeta_\pm$.  In general, due to
the coupling of the form $(\psi_+^{*2}\psi_-^2 + h.c.)$ the GL
equations do not have spherical symmetry.  However, within the
assumption of small $d_{xy}$-wave component, we can apply a partial
wave expansion for $\psi_-$:
\begin{equation}
\psi_- = \sum_n f_-^{(n)}(r)e^{in\theta}.
\end{equation}
It is straightforward to show that the only non-vanishing component
is of $n=1$, and that the asymptotic results given above have no
change.

We also provide the numerical results for the distribution of the
order parameter and the magnetic induction in Fig. \ref{fig:SV}.
As shown in the Fig.  \ref{fig:SV} (a), the magnitude $f_-(r)$ of the
$d_{xy}$-wave component increases as the $d_{x^2-y^2}$-$d_{xy}$
coupling strength $\nu_+$ ($\nu_-$) increases and as the temperature
$T$ decreases.  However, $f_-(r)$ is so small that its effect on
$f_+(r)$ and $B(r)$ is negligible; the isolated single vortex
structure of this model is very similar to that of the conventional
$s$-wave GL theory.

\section{Vortex Lattice and Magnetization near the Upper Critical Field}

In the vicinity of the upper critical field $H_{c2}$, the magnitudes
of the order parameters are small and thus the non-linear terms in the
GL equations Eqs. (\ref{eq:GL1},\ref{eq:GL2}) are negligible.  The
magnetic field are also assumed to be constant all over the space since
the inter-vortex spacing is much less than the penetration depth
$\lambda_+$.  Then the GL equations are reduced to a linearized form:
\begin{eqnarray}
(\Pi^2-1)\psi_+  - \nu_+B_z\psi_-  & = &  0, 
\label{eq:LGLa} \\
(\Pi^2 + \Xi)\psi_- - \nu_-B_z\psi_+ & = &  0.
\label{eq:LGLb}
\end{eqnarray}
Each of the solutions $\psi_{L\pm}$ is just a linear combination of
the wave functions in the lowest Landau level, which is infinitely
degenerate, and determined so as to minimize the GL free energy.

We minimize the free energy by generalizing the Abrikosov's
procedures \cite{Abriko57,deGenn89}.  In these procedures, 
the two components of the order parameter satisfy
\begin{equation}
\frac{\psi_{L-}}{\psi_{L+}} 
= \frac{1}{\nu_+}\frac{H_{c2}-1}{H_{c2}},
\end{equation}
and that the current contribution is given by 
\begin{equation}
\kappa^2J_L 
= -\frac{1}{2}\nabla\times\Bigl\{
  |\psi_{L+}|^2 + \mu^{-1}|\psi_{L-}|^2 
  - \nu_+(\psi_{L+}^*\psi_{L-} + h.c.)
\Bigr\}\hat{z}.
\end{equation}
Then the magnetization\cite{endnote32} and the GL free energy are given by 
\begin{equation}
4\pi M = -\frac{H_{c2}-H}{(2\kappa_{\rm eff}^2-1)\beta_A},
\label{eq:M}
\end{equation}
\begin{equation}
G_s(H) 
= G_n(H_{c2})
+ 2\kappa^2\left[(H_{c2}^2 - H^2)
-\frac{(H_{c2}-H)^2}{(2\kappa_{\rm eff}^2-1)\beta_A}
\right]
\label{eq:GLenergy}
\end{equation}
where  
$\beta_A\equiv\langle\:f_+^4\:\rangle/\langle\:f_+^2\rangle^2$
 is a still undetermined parameter, 
and $\kappa_{\rm eff}^2(T) = \kappa^2\;G(T)/F^2(T)$ with 
\begin{eqnarray*}
R(T)   & = &  \frac{1}{\nu_+}\;\frac{H_{c2}(T)-1}{H_{c2}(T)} \\
F(T)   & = & 1 - 2\nu_+ R + \mu^{-1}R^2 \\
G(T)   & = & 1 + 2(\chi_+ + \zeta_+)R^2 + \mu^{-1}\Upsilon R^4.
\end{eqnarray*}
The Abrikosov parameter $\beta_A$ concerns the vortex lattice
structure and is determined by minimizing the GL free energy.  Since
$\psi_{L+}$ has the same form as the $s$-wave Abrikosov solution, it
is obvious that the free energy is minimized for the triangular vortex
lattice ($\beta_A =1.16$)\cite{Kleine64}.  The magnetization in Eq.
(\ref{eq:M}) shows the usual form of the conventional $s$-wave GL
theory\cite{Abriko57,deGenn89}, but the slope $4\pi dM/dH$ depends on
the temperature.

\section{Discussions}

There is no direct observation of the single vortex structure in a
HTSC, while Keimer {\it et al.}\cite{Keimer94} and Maggio-Aprile
{\it et al.}\cite{Maggio95} observed oblique lattices in their SANS
and STM experiments on YBCO compound, respectively.  Their result of
oblique lattice is not consistent with our result of triangular lattice.
Both the single vortex structure and the vortex lattice structure is
also different from theoretical predictions based on the
$d_{x^2-y^2}$-$s$ admixture
scenario.\cite{Berlin95,Franzx96,Xuxxxx95,Xuxxxx96,Wonxxx95}

The single vortex structure can have a significant influence on the
structure of the vortex lattice in many-vortex problem.   It is quite
interesting to note that the four-fold symmetry of $d$-$s$
admixed vortex is an implication of the mixed gradient coupling in its
model.  In our model, there is no mixed gradient term up to quadratic
order.  Very recently, Ichioka {\it et al.}\cite{Ichiok96} argued that
a non-local correction is required and associated GL theory should
enclose the fourth order mixed gradient term.  They observed that the
$d_{x^2-y^2}$-wave component has the four-lobe shape and $d_{xy}$-wave
component the shape of octofoil.  Although in this present paper we
put emphasis on the effect of mixing through the direct coupling to
the magnetic field, the result of Ichioka {\it et al.} strongly
suggests that the $d_{x^2-y^2}$-$d_{xy}$ admixture can leads to an oblique
vortex lattice.  Furthermore, even in the present GL model, for the
intermediate field regime off the extreme regions near to $H_{c1}$ or
$H_{c2}$, the order parameter distribution is not spherically
symmetric.  Therefore further study is required for that region,
because the non-spherical symmetry of the supercurrent
distribution around a vortex is not compatible with  the triangular
lattice symmetry.

The strong temperature dependence of $\kappa_{\rm eff}(T)$ in Eq.
(\ref{eq:M}) reminiscent of $\kappa_{2}(T)$ in the microscopic
consideration of conventional superconductors by Maki\cite{Makixx65}
and Eilenberger\cite{Eilenb67}.  Roughly speaking, the difference
between $\kappa_2(T)$ and $\kappa$ for that case was due to the
nonlocality of the electromagnetic response of the superconductors.
In our case, the strong $T$-dependence of $\kappa_{\rm eff}(T)$ has an
interesting interpretation.  In contrast to the low
field limit, $\psi_-$ is order of $\psi_+$ at high fields ($H\sim
H_{c2})$ and low temperatures for $\nu_p \sim 1$, and the associated
paramagnetic current significantly affects the magnetization curve
$4\pi M(H)$ in low temperature region.  Several experiments reported
the strong temperature dependence of the slope $4\pi dM/dH$ near
$H_{c2}(T)$ in HTSC.\cite{Koyama96}

\section{Conclusions}
We investigated a GL theory for vortex structures and magnetization in
a $d$-wave superconductor.  In the GL theory, we assumed the
$d_{x^2-y^2}$-wave symmetry of the superconducting ground state, and
the admixture of $d_{x^2-y^2}$-wave and $d_{xy}$ symmetry in the
presence of the vortices.   The structure of an isolated single vortex
was studied asymptotically and numerically at low field region ($H\sim
H_{c1}$).  The isolated single vortex is similar to conventional
$s$-wave vortex, and has almost spherically symmetric supercurrent
distribution around it.   The vortex lattice structure and
magnetization were studied analytically at high fields near the upper
critical field $H_{c2}$.  The vortices arrange in a triangular
lattice, and the magnetization curve $4\pi M(H)$ shows a strong
temperature dependence for $\nu_p\sim 1$ due to the paramagnetic
current effect.  Some physical implications of the results were
discussed.  The results were also compared with the experimental
observations and with those of $d$-$s$ scenario.  It was recognized
that further study in the intermediate field region is valuable. 

\section*{Acknowledgments}
We thank T. Koyama and M. Tachiki very much for valuable discussions
and sending us preprints.  This work was supported by the Seoam
Scholarship Foundation, and by the Basic Science Research Institute
Program, Ministry of Education (BSRI-95-2437).

%
% References
%

%
%   TABLE
%
\begin{table}
\caption{Characteristic lengths, scales of physical quantities, and
phenomenological parameters in the GL theory in question.}
\label{table:parameters}

\begin{tabular}{ll}
Characteristic lengths
    & $\displaystyle\xi_+^2 = \frac{\hbar^2}{4m_+|\alpha_+|}$, 
    $\displaystyle\xi_-^2 = \frac{\hbar^2}{4m_-\alpha_-}$, 
    $\displaystyle \lambda_+^2 = \frac{m_+c^2}{8\pi e^2|\Psi_o|^2}$,
     \\
\hline
Characteristic field\tablenote{
Here $H_{c2}^o$ is not the true upper critical field.
The upper critical field can be substantially enhanced in this model.
}
    & $H_{c2}^o = \phi_o/2\pi\xi_+^2$
    \\
\hline
Fundamental parameters
    & $\Psi_o^2 = |\alpha_+|/\beta_+$,
    $\Phi_o/2\pi = \hbar c/2e$, 
    $\kappa = \lambda_+/\xi_+$  \\
\hline
Auxiliary parameters
    & $\Xi = \xi_+^2/\xi_-^2$,
    $\Upsilon = m_-\beta_-/m_+\beta_+$, 
    $\mu = m_-/m_+$,\\
    & 
    $\nu_\pm = \gamma_p m_\pm c/e\hbar$,
    $\nu_p^2 = \nu_+\nu_-$, \\
    & 
    $\chi_+ = \beta_X/\beta$, $\chi_- = \mu\chi_+$  \\
    & 
    $\zeta_+ = \beta_Y/\beta$, $\zeta_- = \mu\zeta_+$  \\
\hline
Reduced units
    & $\Psi_\pm/\Psi_o = \psi_\pm$, 
        $T/T_c\to T$, ${\bf r}/\xi_+ \to {\bf r}$, \\
    & ${\bf B}/H_{c2}^o \to {\bf B}$, 
    ${\bf A}/\xi_+H_{c2}^o \to {\bf A}$, 
    ${\bf J}/(cH_{c2}^o/4\pi\xi_+) \to {\bf J}$
\end{tabular}

\end{table}

%
%   FIGURE CAPTIONS
%
\begin{figure}
%\epsfclipon\epsfverbosetrue\
%\epsfxsize=7cm
%\epsfbox{mschoi-fig1.eps}

\caption{Distribution of the magnitudes of the order parameter and the
magnetic induction associated with an isolated single vortex, for
different values of $T$ and $\nu_+$: $T=0.1, 0.9$ and $\nu_+=0.1,
0.9$.  Other parameters are set as $\mu = 1.0$, $\Xi(0)=1.0$,
$\Upsilon = 1.0$, $\chi_+=0.8$.  }
\label{fig:SV}
\end{figure}

\begin{figure}
%\epsfclipon\epsfverbosetrue\
%\epsfxsize=7cm
%\epsfbox{mschoi-fig2.eps}

\caption{Temperature dependence of the $\kappa_{\rm eff}^2$ for three
different values of $\nu_p$.  Other parameters are set as  $\mu =
1.0$, $\Xi(0)=1.0$, $\Upsilon = 1.0$, $\chi_+=0.8$.}
\label{fig:kappa}
\end{figure}


\begin{thebibliography}{10}
\bibitem{endnote11}
We neglect the possibility of acidental degeneracy.
\bibitem{endnote33}
In the work of Koyama and Tachiki, the GL free energy functional was derived
  only upto quadractic order.
\bibitem{endnote21}
The inter-vortex distance is much larger than the penetration depth.
\bibitem{endnote1}
Precisely speaking, the necessary condistion is $\varphi(r,\theta) = n\theta +
  C$ (with $n$ integer). But the lowest energy states correspond to $n=\pm 1$,
  and we take $-1$ just for a convenience.
\bibitem{endnote31}
$C_0$ is related to $B_0$, and determined by going higher order.
\end{thebibliography}
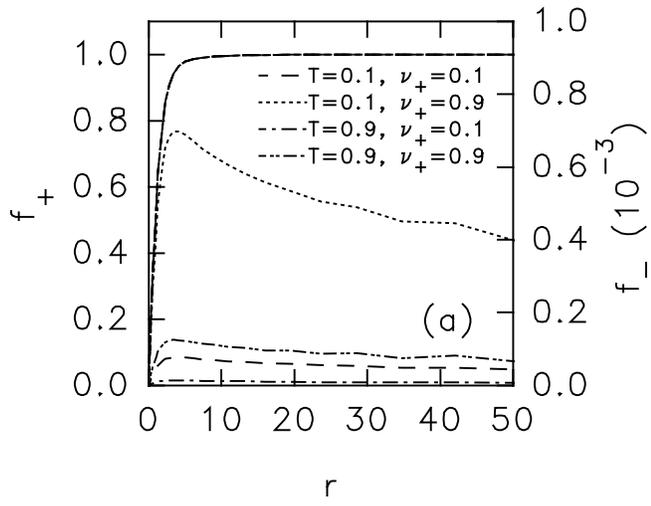
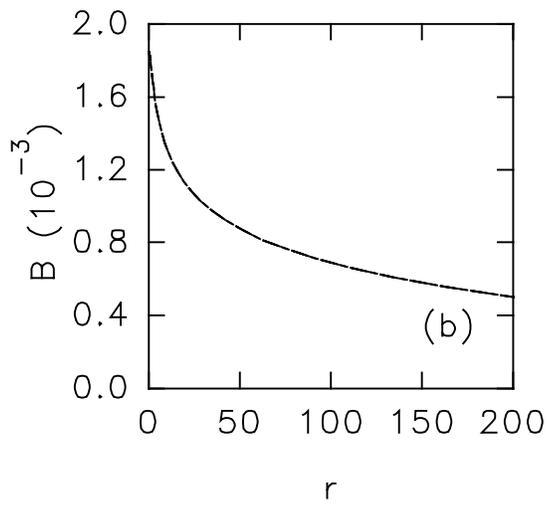
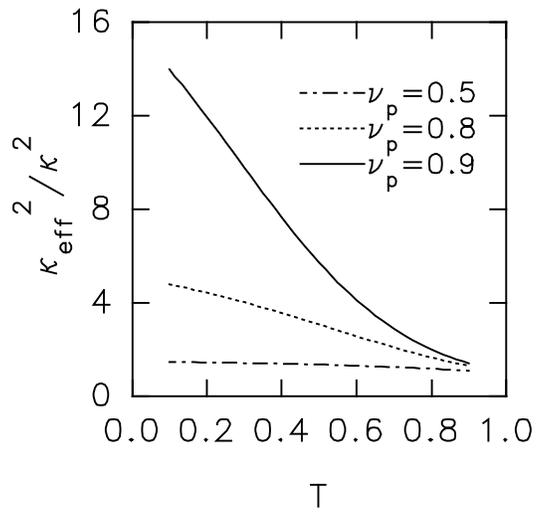
\end{document}